\documentclass[aps,prl,12pt,oneside,onecolumn]{revtex4}

\usepackage{graphicx}
\usepackage{dcolumn}
\usepackage{epstopdf}
\usepackage{bm}
\usepackage{amssymb}
\usepackage{amsmath}
\usepackage{parskip}
\usepackage{multirow}

\usepackage{subfigure}
\usepackage[sort&compress]{natbib}
\newcommand{\degree}{\ensuremath{^\circ}}

\begin{document}

\bibliographystyle{naturemag}

\title{Novel Itinerant Antiferromagnet TiAu}

\author{E. Svanidze,$^1$ Jiakui K. Wang,$^1$ T. Besara,$^2$ L. Liu,$^3$ Q. Huang,$^4$ T. Siegrist,$^2$ B. Frandsen,$^3$ J. W. Lynn,$^4$ Andriy H. Nevidomskyy,$^1$ Monika Barbara Gam\.za,$^{5,6}$ M. C. Aronson,$^{5,7}$ Y. J. Uemura$^3$ and E. Morosan$^1$}

\address{$^1$ Department of Physics and Astronomy, Rice University, Houston, TX, 77005 USA}
\address{$^2$ National High Magnetic Field Laboratory, Florida State University, Tallahassee, FL, 32306 USA}
\address{$^3$ Department of Physics, Columbia University, New York, New York 10027, USA}
\address{$^4$ NIST Center for Neutron Research, National Institute of Standards and Technology, MD 20899, USA}
\address{$^5$ Condensed Matter Physics and Materials Science Department, Brookhaven National Laboratory, Upton, NY, 11973 USA}
\address{$^6$ Department of Physics, Royal Holloway, University of London, Egham, TW20 0EX, UK}
\address{$^7$ Department of Physics and Astronomy, Stony Brook University, Stony Brook, NY, 11794 USA}


\maketitle
\textbf{
Itinerant and local moment magnetism have substantively different origins, and require distinct theoretical treatment. A unified theory of magnetism has long been sought after, and remains elusive, mainly due to the limited number of known itinerant magnetic systems. In the case of the two such examples discovered several decades ago, the itinerant ferromagnets ZrZn$_2$ \cite{Matthias_1958} and Sc$_3$In\cite{Matthias_1961}, the understanding of their magnetic ground states draws on the existence of \textsl{3d} electrons subject to strong spin fluctuations. Similarly, in Cr\cite{Fawcett_1988}, an elemental itinerant antiferromagnet (IAFM) with a spin density wave (SDW) ground state, its \textsl{3d} character has been deemed crucial to it being magnetic. Here we report the discovery of the first IAFM compound with no magnetic constituents, TiAu. Antiferromagnetic order occurs below a N\'{e}el temperature T$_N~\simeq$ 36 K, about an order of magnitude smaller than in Cr, rendering the spin fluctuations in TiAu more important at low temperatures. This new IAFM challenges the currently limited understanding of weak itinerant antiferromagnetism, while providing long sought-after insights into the effects of spin fluctuations in itinerant electron systems.}

TiAu has been reported to form in three distinct crystal structures, cubic \textit{Pm$\bar{3}$m} \cite{Donkersloot_1970}, tetragonal \textit{P4/nmm} \cite{Schubert_1962} and orthorhombic \textit{Pmma} \cite{Donkersloot_1970}, posing an inherent difficulty in synthesizing it as a single phase. It comes as no surprise then, that no reports of physical properties of TiAu exist. Here we report the magnetic and electronic properties of phase pure orthorhombic TiAu, and show that it is a new itinerant electron antiferromagnet.

\begin{figure}[b!]
\centering
\includegraphics[width=\columnwidth]{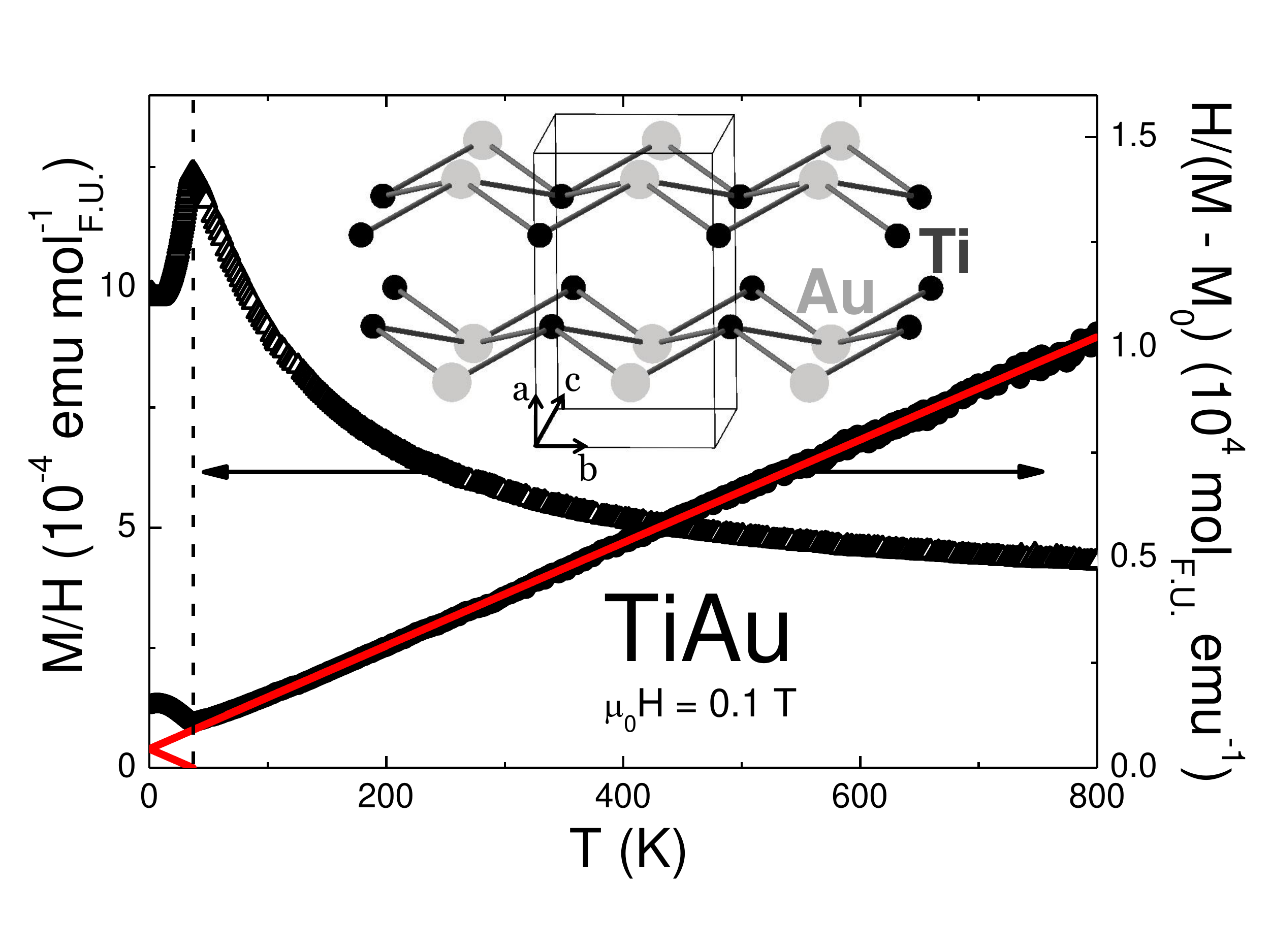}
\caption{\textbf{Temperature-dependent magnetization of TiAu.} Left axis: zero-field cooled magnetic susceptibility as a function of temperature for $\mu_0 H$ = 0.1 T applied field. Right axis: inverse susceptibility $H/M$ along with a Curie-Weiss-\textit{like} fit (solid line), with $\theta$ $\approx$ - 37 K. Inset: the crystal structure of TiAu with Ti (small) and Au (large) atoms. (1 emu = 10 A/cm$^2$)}
\label{MT}
\end{figure}

The first evidence for the antiferromagnetic ground state is the cusp around 36 K in the temperature-dependent magnetic susceptibility $M(T)/H$, shown in Fig. \ref{MT} (left axis). By analogy with local moment antiferromagnets (AFMs), the TiAu zero-field-cooled and field-cooled data are indistinguishable. The value of the temperature-independent Pauli susceptibility, calculated from the magnetic density of states $\chi_0\approx$ 0.2$\cdot$10$^{-3}$ emu mol$^{-1}$, agrees well with the experimental one M$_0/$H $\approx$ 0.3$\cdot$10$^{-3}$ emu mol$^{-1}$. Upon warming above the N\'{e}el temperature, the inverse susceptibility $H/(M-M_0)$ (right axis, Fig. \ref{MT}) is linear in temperature up to 800 K. Such linear T dependence of H/M has been long considered the hallmark of local moment magnetism, until it was observed in the weak itinerant ferromagnets (IFMs) without local moments, ZrZn$_2$ \cite{Matthias_1958} and Sc$_3$In \cite{Matthias_1961}. Puzzling at first, the origin of this behavior was reconciled in the case of IFMs, when spin fluctuation effects were considered by Moriya \cite{Moriya_1973, Moriya_1973b}. The self-consistent renormalization theory unified the local and itinerant pictures of ferromagnetism, and postulated a new origin for the Curie-Weiss-\textit{like} susceptibility in the latter, as the interactions of the spatially extended modes of spin fluctuations \cite{Moriya_1985, Hasegawa_1974}. TiAu however is an IAFM, and no existing theory accounts for an itinerant antiferromagnetic ground state if neither Ti nor Au have conventional local moments. XPS analysis suggests that Ti is close to the non-magnetic $4+$ oxidation state (see Supplementary Material). This is striking in light of the high magnetic volume fraction observed in the muon spin relaxation measurements presented below, which, together with the single phase neutron and XPS patterns in Figs. SM1 and SM2 in the Supplementary Material, indicates that the observed magnetic behavior is indeed intrinsic. Therefore \textit{Pmma} TiAu is \textit{the first} IAFM compound with no magnetic constituents, with the magnetic ground state strongly affected by spin fluctuations.

\begin{figure}[t!]
\centering
\includegraphics[width=0.8\columnwidth]{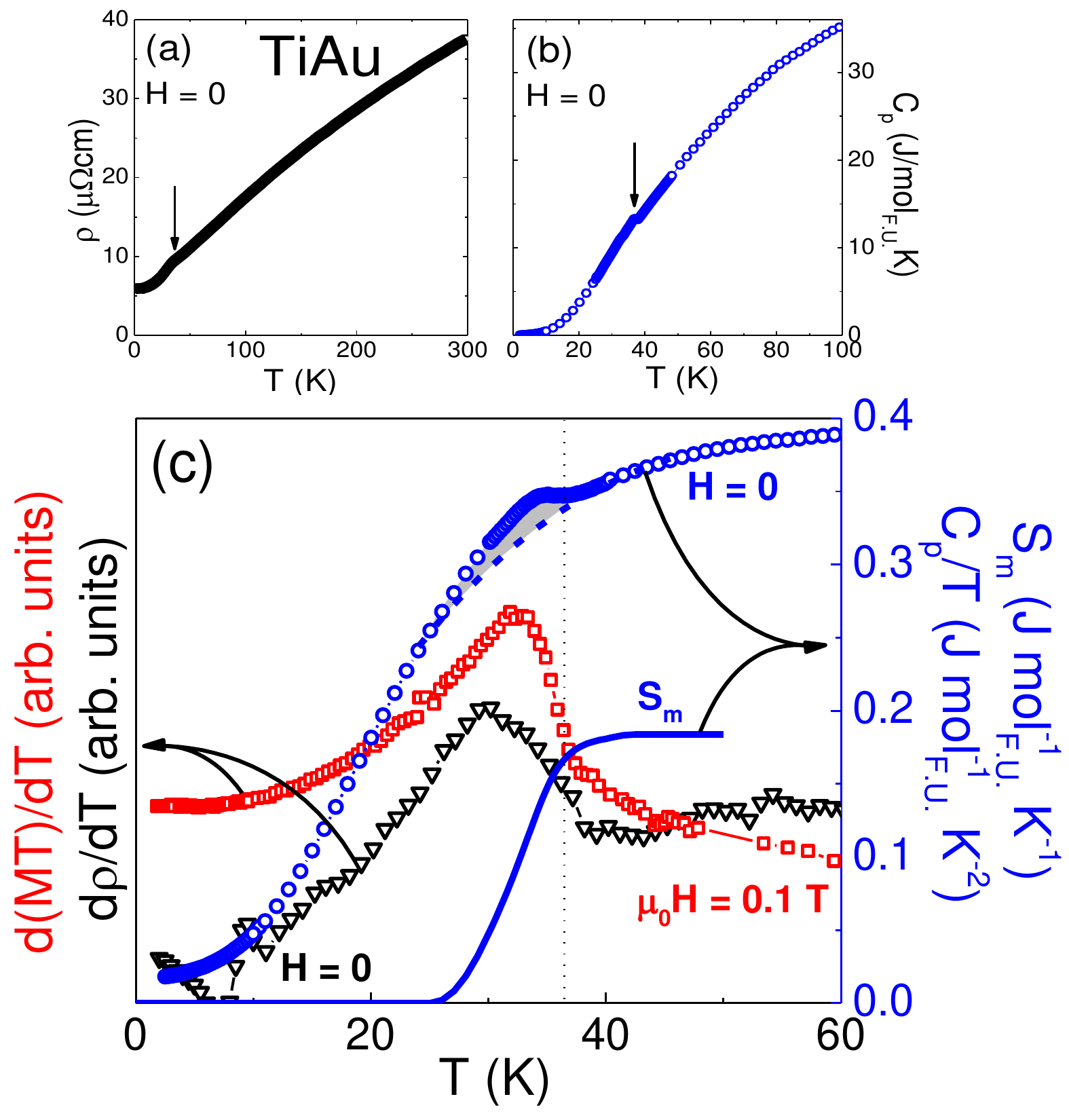}
\caption{\textbf{Specific heat and resistivity of TiAu.} $H$ = 0 temperature-dependent resistivity (a) and specific heat (b). (c) The ordering temperature T$_N$ (vertical dotted line) for TiAu determined from peaks in the temperature derivatives of resistivity, $d\rho/dT$ (black triangles), and $MT$, $d(MT)/dT$ (red squares), and in $C_p/T$ (blue circles). The entropy $S_m$ (solid blue line, right axis) is calculated by subtracting a polynomial non-magnetic component (dashed line) from the measured specific heat data.}
\label{RTCT}
\end{figure}

Remarkable for a weak itinerant (ferro- or antiferro-) magnet, the electrical resistivity (Fig. \ref{RTCT}a) and specific heat data (Fig. \ref{RTCT}b) also show signatures of the phase transition around 36 K. The $H = 0$ $\rho(T)$ data are typical of a good metal, decreasing nearly linearly from room temperature down to $\sim$ 40 K. A drop of about 10\%, similar to the loss of spin-disorder scattering, occurs at T$_N$ (Fig. \ref{RTCT}a). Although often the gap opening associated with the SDW ordering results in a resistivity increase, a similar drop was observed in BaFe$_2$As$_2$ single crystals \cite{Wang_2009}. In the absence of local moment ordering, the decrease in the resistivity at T$_N$ results from the balance of the loss of scattering due to Fermi surface nesting (see below) and the gap opening due to the SDW AFM state. At the same temperature in TiAu, a small peak becomes visible in the specific heat data $C_p$ (Fig. \ref{RTCT}b), such that $T_N$ in this AFM metal can be determined, as shown by Fisher \cite{Fisher_1962, Fisher_1968}, from peaks in $C_p$ (most visible in $C_p$/T), $d(MT)/dT$ and $d\rho/dT$ (Fig. \ref{RTCT}c). Distinguishing between local and itinerant moment magnetism is inherently difficult, especially in the nearly unexplored realm of IAFMs. It is therefore striking that in TiAu, abundant evidence points towards its itinerant magnetic moment character. The fact that the peak in $C_p$ is not as strong as the Fisher prediction \cite{Fisher_1962} is one such argument favoring the itinerant moment scenario in TiAu. Another argument is the small magnetic entropy S$_m$ (grey area, Fig. \ref{RTCT}c) associated with the transition (solid blue line, Fig. \ref{RTCT}c). Even though the S$_m$ calculated after assuming a polynomial non-magnetic C$_p$ around the transition (dashed line, Fig. \ref{RTCT}c) is an underestimate, it amounts to only 0.2 J mol$^{-1}$ K$^{-1}$ or $\sim$3$\%$ $R$ln$2$.

\begin{figure}[t!]
\centering
\includegraphics[width=0.8\columnwidth]{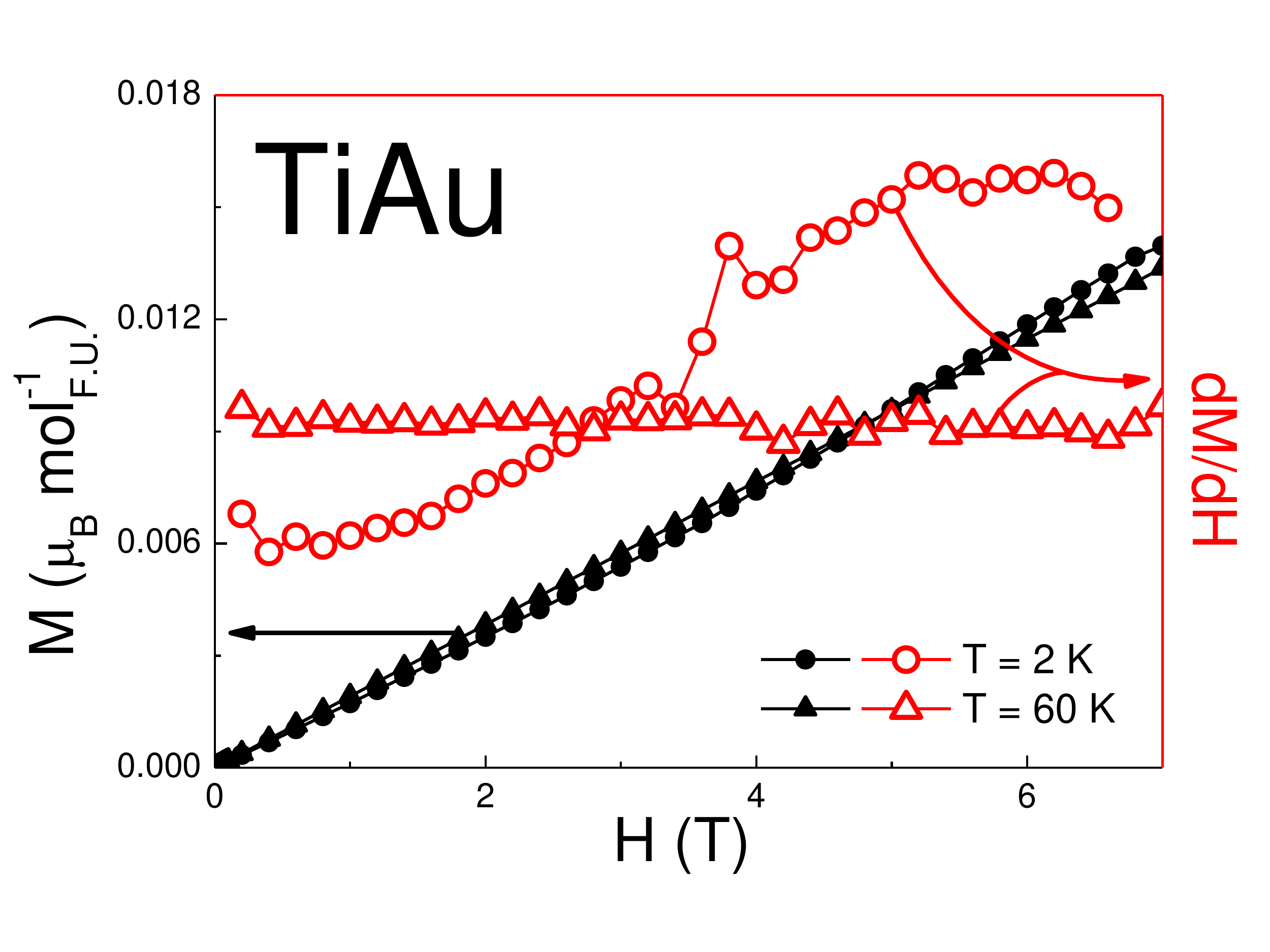}
\caption{\textbf{Field-dependent magnetization of TiAu.} The magnetization isotherms $M(H)$ (solid, left axis) and the derivative $dM/dH$ (open, right axis) for $T$ = 2 K (circles) and 60 K (triangles). No saturation is achieved for magnetic fields up to 7 T. A metamagnetic transition is observed around 4 T in the $T$ = 2 K isotherm, but not in the one above the magnetic order.}
\label{MH}
\end{figure}

Despite the remarkably large paramagnetic moment $\mu_{PM} \simeq~0.8~\mu_B$, derived from the Curie-Weiss-\textit{like} fit of the inverse susceptibility (Fig. \ref{MT}a, right axis), the field-dependent magnetization $M(H)$ does not saturate up to 7 T, and the maximum measured magnetization is only 0.01 $\mu_B$ (Fig. \ref{MH}). A closer look at the low temperature $M(H)$ reveals a weak  metamagnetic transition starting around $\mu_0 H = 3.6$ T for $T = 2$ K (circles, Fig. \ref{MH}). This is most apparent in the derivative $dM/dH$ (open symbols) rather than in the as-measured isotherms (full symbols), with the latter nearly indistinguishable well below ($T = 2$ K) and above ($T = 60$ K) the magnetic ordering temperature. It has been shown by Sandeman \textit{et. al.} \cite{Sandeman_2003} that, within the Stoner theory, the presence of a sharp double peak structure in the electronic DOS sufficiently close to the Fermi level results in a metamagnetic transition. The argument requires that the paramagnetic Fermi level lie in between the two peaks of the DOS, and this is indeed revealed by the band structure calculation for TiAu, shown in the inset of Fig. \ref{DOS}a below. It results that, as the Fermi sea is polarized by the applied magnetic field $H$, the majority and minority spin Fermi levels feel the effect of the two DOS peaks at different values of induced magnetization. The DOS peak that is closest to the Fermi level will lead to a sharp increase (decrease) in the population of the majority (minority) spin band, resulting in a metamagnetic transition.

\begin{figure}[b!]
\centering
\includegraphics[width=0.8\columnwidth]{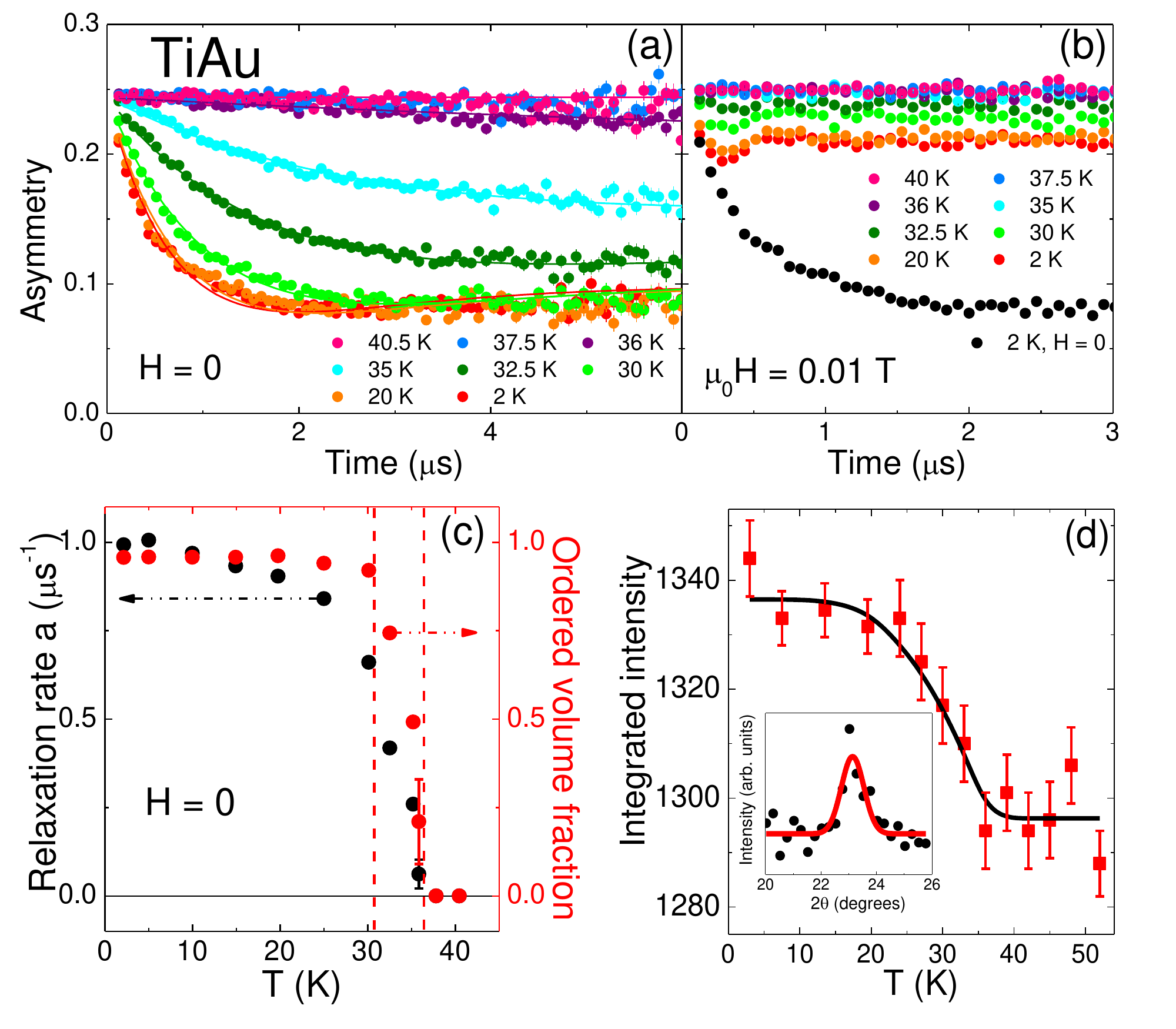}
\caption{\textbf{Muon spin relaxation and neutron diffraction.} (a) Time dependence of the asymmetry, fit with Eq. 1 (solid lines). (b) A small applied longitudinal field $\mu_0H = 0.01$ T eliminates the relaxation response. (c) Relaxation rate $a$ (triangles, left) and volume fraction (circles, right) as a function temperature. (d) Integrated intensity of the (0, $\pi$/$b$, 0) TiAu magnetic Bragg peak as a function of temperature with mean-field fit ($T_N$ = 36(2), black curve). Inset: the (0, $\pi$/$b$, 0) magnetic peak fit with a resolution-limited Gaussian (red line). Uncertainties are statistical in origin and represent one standard deviation.}
\label{muSR}
\end{figure}

\begin{figure}[b!]
\centering
\includegraphics[width=1\columnwidth]{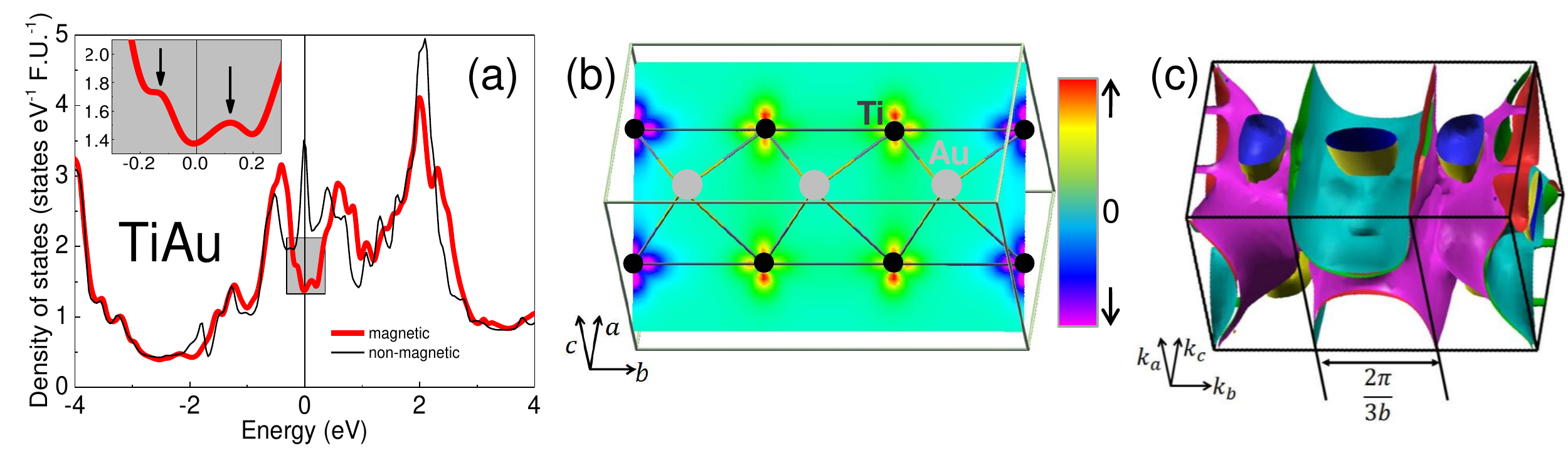}
\caption{\textbf{Band structure calculations for TiAu.} (a) The non-magnetic density of states (thin black line) exhibits a peak close to the Fermi surface, similar to that seen in other itinerant magnets. For the AFM1 ground state, the finite total density of states (thick red line) at the Fermi energy is flanked by two peaks around 0.1 eV (inset), which explains the metamagnetic transition at low $T$ (see text). (b) The electron spin density shows a modulation along the b axis, consistent with the $k$ = 2$\pi/$(3b) nesting shown in (c). Fermi surface with nesting vector $Q_{calc} = (0, 2\pi/3b, 0)$ is shown in (c). Separated Fermi surface plot is available in the Supplementary Material.}
\label{DOS}
\end{figure}

Muon spin relaxation ($\mu$SR) data shown in Fig. \ref{muSR} unambiguously confirm the static magnetic order developing in the full volume fraction, with the transition temperature corresponding to the anomaly in the magnetic susceptibility, resistivity and specific heat shown in Fig. \ref{RTCT}c. For temperatures above 35 K, the total asymmetry undergoes a negligibly small relaxation, signaling lack of static magnetic order (Fig. \ref{muSR}a). In the time spectra observed in zero field (ZF), a fast decaying front end begins to develop around 35 K, and becomes more pronounced for lower temperatures. This early time decay results from the build up of static internal field, since a small longitudinal field (LF), $\mu_0H = 0.01$ T, eliminates this relaxation via the decoupling effect (Fig. \ref{muSR}b). The time spectra in ZF are fitted with the relaxation function, expected for a Lorentzian distribution of local fields \cite{Uemura_1985}:

\begin{equation}
G(t) = f \Big(\frac{1}{3} + \frac{2}{3}(1-at)e^{-at}\Big) + (1 - f)
\end{equation}

\noindent where $f$ represents the volume fraction with static magnetic order. As shown in the Supplementary Material, a slightly better fit can be obtained by a phenomenological function composed of the sum of “two exponential” terms; but current analysis is more applicable since it can be directly compared with $\mu$SR results in other systems, as shown in Table SM1 in Supplementary Material. The temperature dependence of the relaxation rate $a$ and the magnetic volume fraction $f$ are shown in Fig. \ref{muSR}c. A reasonably sharp transition occurs below T$_N$ = 36 K to a state with 100$\%$ ordered volume, preceeded on cooling by a small temperature region around T$_N$ characterized by the finite volume fraction $f$, which suggests co-existence of ordered and paramagnetic volumes in real space \textit{via} phase separation.

A remarkable feature found in both zero field (ZF) and longitudinal field (LF) time spectra is the absence of dynamic relaxation, expected for critical slowing down of spin fluctuations around T$_N$. Such an effect should have resulted in the $1/T_1$ relaxation of the asymmetry measured in $\mu_0H = 0.01$ T, since this LF can eliminate the effect of static magnetism, while dynamic effects survive in a small LF.  Together with the ZF relaxation function (Eq. 1) which solely involves static effects, the present data indicate complete absence of dynamic critical behavior.  Although occuring in a limited temperature region, the aforementioned phase separation indicates that the transition is likely first-order, without dynamic critical behavior. As shown in Table SM1 (Supplementary Material), similar absence of dynamic critical behavior associated with phase separation was observed in $\mu$SR studies of the itinerant helimagnet MnSi in an applied pressure of 13 - 15 kbar \cite{Uemura_2007}, near the pressure-tuned quantum crossover to the paramagnetic phase.  Such tendencies were also seen in the itinerant ferromagnet (Sr,Ca)RuO$_3$ close to the disappearance of static magnetic order around a Ca concentration of 0.7 \cite{Gat_2011}. The first-order transition may be a generic feature of weak magnetic order in itinerant-electron systems \cite{Belitz_1999}.

As Table SM1 (Supplementary Material) shows, the magnitude of the internal magnetic field in TiAu in the ordered state is remarkably small, compared to $\mu$SR results in other itinerant electron systems, dilute alloy spin glasses or the incommensurate SDW system (Sr$_{1.5}$Ca$_{0.5}$)RuO$_4$ \cite{Carlo_2012}. Although this indicates a very small ordered moment in TiAu, it is not possible to estimate the moment size since the hyperfine coupling constant could depend strongly on the assumption of the location of muon sites. The lineshape of the ZF $\mu$SR data in Eq. (1) is obtained for the case of dilute-alloy spin glasses where the local field at the muon site varies due to different distances to the moment site \cite{Uemura_1985}. However, the same lineshape was also observed in (Sr$_{1.5}$Ca$_{0.5}$)RuO$_4$ in which incommensurate SDW order was recently confirmed by neutron scattering. Therefore, it is difficult to determine spin structure of the system from the present data alone. 

Neutron diffraction measurements above (T = 60 K) and below (T = 2 K) the ordering temperature (Supplementary Material) show virtually indistinguishable patterns. However, a magnetic peak (inset, Fig. \ref{muSR}d) in the low temperature data is revealed by the difference between the two measurements, consistent with long range magnetic order and a small itinerant moment ordering at $T_N$ = 36(2) K. The magnitude of the itinerant moment is estimated to be 0.15 $\mu_B$/Ti, assuming a spin direction perpendicular to the $b$ axis. Together with the present $\mu$SR data, that show a magnetic phase fraction of 100$\%$, these results  eliminate the possibility that the observed magnetism is due to dilute magnetic impurities or a minority phase. These arguments demonstrate that the magnetism of the present system is due to a generic feature of TiAu.


Of particular interest is the comparison between the experimental evidence for the antiferromagnetic order in TiAu with the theoretical results from band structure calculations. A number of possible magnetic configurations were considered: ferromagnetic (FM), AFM SDW with modulation vectors $Q1 = (0, 2\pi/3b, 0)$ (AFM$1$) and $Q2 = (0, \pi/b, 0)$ (AFM$2$). Their energies relative to the non-magnetic state were estimated to be $E_{\text{FM}} = - 25$ meV/Ti, $E_{\text{AFM1}} = - 42$ meV/Ti and $E_{\text{AFM2}} = - 30$ meV/Ti, respectively. These energy values suggest that AFM1 would be the ground state configuration, while the neutron diffraction data show a wavevector $Q_{exp}$ = $(0, \pi/b, 0)$ (Fig. \ref{muSR}d), which corresponds to the AFM2 state. We note that the calculated energy difference between the AFM1 and AFM2 states is small, $\approx$ 12 meV/Ti, comparable to the accuracy threshold of the spin-polarized DFT calculation. Furthermore, the calculation yields a small ordered magnetic moment $\mu_{calc}$ for all surveyed configurations, $0.52~\mu_B$/Ti $\leq \mu_{calc} \leq 0.74~\mu_B$/Ti, reflecting the itinerant nature of the ordered moment. Given that neutron scattering measurements estimate the ordered magnetic moment in TiAu to be 0.15 $\mu_B$/Ti, it appears that the DFT calculation overestimates the moment. This overestimate in TiAu could be explained by strong spin fluctuations which cannot be accounted for by the DFT calculations. The same problem was also encountered in Fe pnictides, which are also itinerant SDW AFM compounds \cite{Cruz_2008, Ma_2008}. In the case of TiAu this may be remedied by future DMFT calculations.

The Fermi surface of the non-magnetic TiAu (Fig. \ref{DOS}c) exhibits four half-cylindrical surfaces with large nearly nested regions in the $k_b$ direction. These are more two dimensional than the nested electron-hole pockets in Cr \cite{Fawcett_1988}, the well-known elemental antiferromagnet. This may explain the considerably larger drop in the relative magnetic susceptibility $\Delta$M$/$M at T$_N$ in TiAu compared to Cr, where $\Delta$M/M = [M$_{T_N}$ - M$_{T=0}]$/M$_{T_N}$. In TiAu, $\Delta$M/M is $\sim$ 20$\%$, nearly five times larger than in Cr \cite{Fawcett_1988}. In the latter, the small magnetization decrease at T$_N$ had been attributed to the small spin susceptibility (and not the larger orbital component) being affected by the gap associated with the SDW transition. Conversely, the larger magnetization change in TiAu might indicate a sizable effect on the orbital magnetization, as the SDW transition is now associated with more two dimensional nesting than that in Cr.

Overall, our data convincingly show that TiAu is a new IAFM compound, the first of its kind, and analogous to the only two IFMs with no magnetic elements, Sc$_3$In and ZrZn$_2$. With ample evidence for the itinerant character of the magnetic state in TiAu, including the small magnetic moment in the ordered state compared to the paramagnetic moment, small magnetic entropy at T$_N$, it is readily apparent that strong spin fluctuations are at play in this novel magnetic system. The exact role of the spin fluctuations, their strength, as well as the details of the magnetic structure in the ordered state, remain to be fully elucidated with further experiments. Additionally, doping experiments are underway \cite{Svanidze_2014} and seem to cast the magnetism in TiAu as robust albeit extremely complex. Ultimately, the search for IAFM materials appears to be a promising avenue for furthering our understanding of the complex magnetism, and providing the unifying picture for local and itinerant moment magnetism.

\vspace{5mm}

\noindent \textbf{METHODS:} Polycrystalline samples of TiAu were prepared by arcmelting, with mass losses no more than 0.3 \%. The hardness of the arcmelted samples rendered powder x-ray diffraction experiments difficult, and therefore x-ray diffraction data were collected at room temperature off the cross-section (about 3 mm in diameter) of cut and polished specimens using a custom 4-circle Huber diffractometer with graphite monochromator and analyzer in non-dispersive geometry, coupled to a Rigaku rotating anode source producing CuK$\alpha$ radiation. X-ray photoemission spectroscopy was performed on the polished surface of the TiAu sample, using an XPS Phi Quantera spectrometer with a monochromatic Al x-ray source and Ar ion sputtering gun, used to cleanse the surface of contaminants. The alignment was checked by comparing the binding energy of the C$1s$ peak to the published one \cite{Nist_2012}. More details can be found in the Supplementary Material.

DC magnetization was measured in a Quantum Design (QD) Magnetic Property Measurement System (MPMS) from 2 K to 400 K. At temperatures above 400 K, the magnetization data were collected using the Vibrating Sample Magnetometer (VSM) option of a QD Physical Property Measurement System (PPMS) equipped with an oven. Specific heat using an adiabatic relaxation method, and four-probe DC resistivity measurements from 2 K to 300 K were carried out in the QD PPMS environment.

Zero-field and longitudinal-field (up to $\mu_0H = 0.01$ T) muon spin relaxation measurements were performed at the M20 channel of TRIUMF, Vancouver, Canada. The cut TiAu samples of thickness around 0.1 cm  and area of about 3.5 cm$^2$ were mounted on a silver foil and oriented perpendicular to the direction of the incoming muons. For all experiments the beam momentum was ~28 MeV/$c$, with muons polarized along the flight direction. 

For the neutron diffraction measurements the sample was sealed with helium exchange gas and mounted in a closed cycle refrigerator with a base temperature of 2.6 K. To search for magnetic scattering the high intensity/coarse resolution BT-7 spectrometer was employed in two-axis mode, with a fixed initial neutron energy of 14.7 meV ($\lambda$ = 2.369 \AA) and collimator (full-width-half-maximum) configuration open - PG(002) monochromator - 80$'$ - sample - 80$'$ radial-collimator - position-sensitive detector \cite{Lynn_2012}. To characterize the sample and search for possible structural changes associated with the magnetic phase transition the BT-1 high resolution powder diffractometer was used. Collimators of 15$'$, 20$'$ and 7$'$ were used before and after the Cu (311) monochromator ($\lambda$ = 1.5401 \AA) and after the sample, respectively, and data were collected in steps of 0.05\degree in the 2$\theta$ range of 3\degree~to 168\degree. The refinements were determined by the Rietveld method with the General Structure Analysis System software, with the results available in the Supplementary Material.

Band structure calculations were performed using the full-potential linearized augmented plane-wave method implemented in the \textit{WIEN2K} package \cite{Wien}. The PBE-GGA was used as the exchange potential, the default generalized gradient approximation for the exchange correlation potential in \textit{WIEN2K} \cite{Perdew_1996}, as described in more detail in the Supplementary Material. The lattice parameters and atomic positions were determined from both neutron and x-ray diffraction and are reported in the Supplementary Material.


\vspace{5mm}

\noindent \textbf{ACKNOWLEDGEMENTS:} We would like to thank M. B. Maple, P. Dai  and M. Foster for useful discussions. The work at Rice was supported by NSF DMR 0847681 (E.M. and E.S.), AFOSR MURI (J.K.W.) and the Welch Foundation grant C-1818 (A.H.N.). Work at Brookhaven National Laboratory (M.B.G. and M.C.A.) was carried out under the auspices of the US Department of Energy, Office of Basic Energy Sciences, under Contract No. DE-AC02-98CH1886. T.B. and T.S. are supported by the U.S. Department of Energy, Office of Basic Sciences under Contract No. DE-SC0008832, the State of Florida and Florida State University. Work at Columbia and TRIUMF (L.L., B.F. and Y.J.U.) is supported by NSF grants DMR-1105961 and OISE-0968226 (PIRE), REIMEI project from JAEA, Japan, and the Friends of Todai Inc. Foundation. The authors thank Z. Deng, T. Munsie, T. Medina and G. Luke for assistance and discussions on $\mu$SR measurements. The identification of any commercial product or trade name does not imply endorsement or recommendation by the National Institute of Standards and Technology.

\vspace{5mm}

\noindent \textbf{AUTHOR CONTRIBUTIONS:} E.M. designed the study, E.S. prepared the samples, performed susceptibility, specific heat, resistivity and x-ray photoemission spectroscopy measurements. E.M and E.S. performed the data analysis and wrote the manuscript with contributions from all authors. J.K.W. and A.N. performed band structure calculations and analysis. T.B. and T.S. were responsible for x-ray measurements and structural characterization. L.L., B.F. and Y.J.U. performed muon spin relaxation measurements, M.B.G. and M.C.A. measured the high-temperature magnetization, J.W.L. and Q.H. carried out neutron diffraction experiments and analysis.

\vspace{5mm}

\noindent \textbf{COMPETING INTERESTS:} The authors declare that they have no competing financial interests.

\vspace{5mm}

\noindent \textbf{CORRESPONDENCE:} Correspondence and requests for materials should be addressed to E.M. at emorosan@rice.edu.

\clearpage

\section{Supplementary Material}

\section{I. X-ray photoemission spectroscopy}

\begin{figure}[b!]
\renewcommand{\figurename}{Fig. SM}
\centering
\includegraphics[width=0.75\columnwidth]{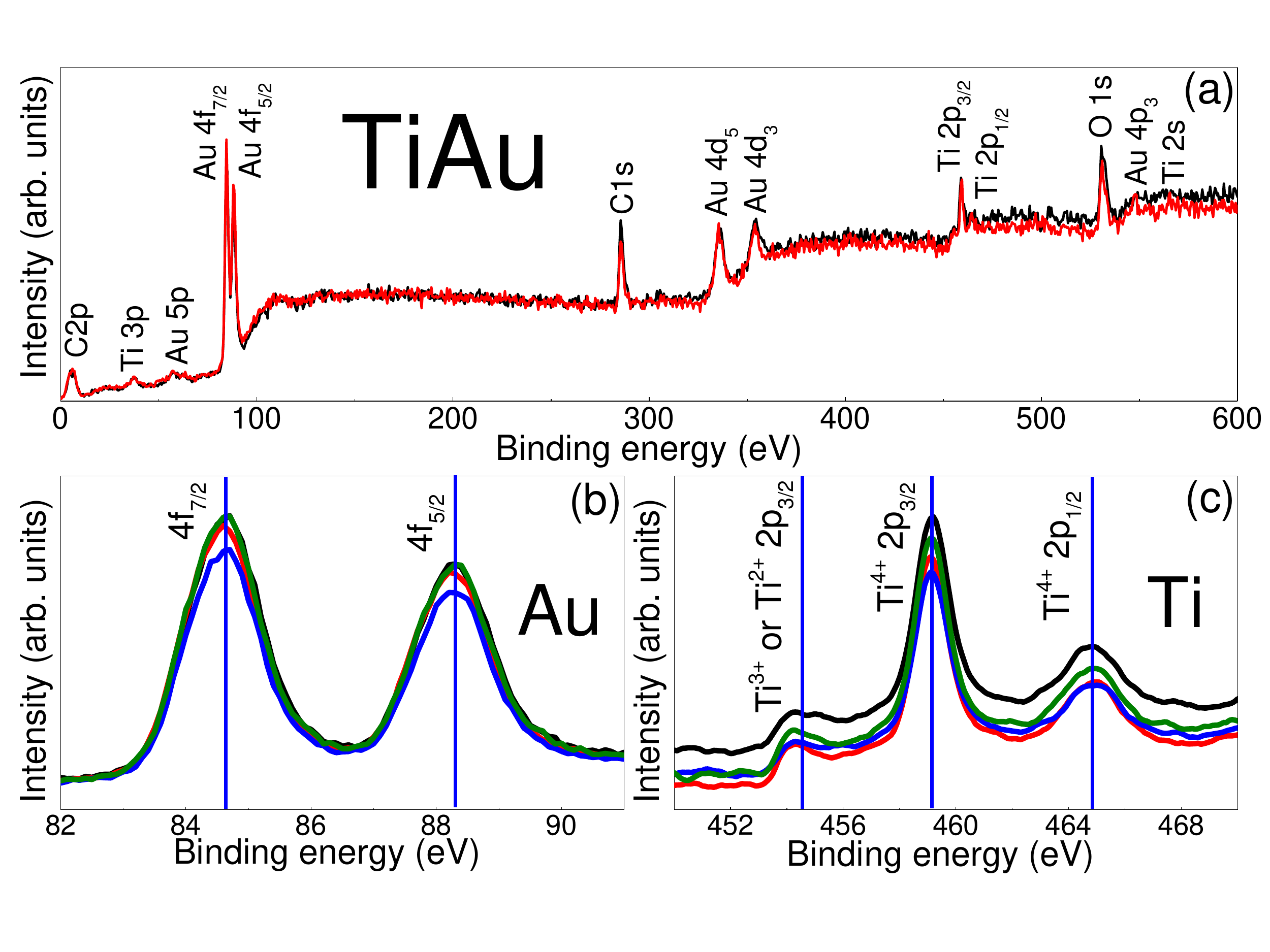}
\caption{\textbf{X-ray photoemission spectroscopy:} (a) survey scan along with elemental scans for Au (b) and Ti (c).}
\label{XPS}
\end{figure}

X-ray photoemission spectroscopy (XPS) is a technique extremely sensitive in resolving the elemental composition\cite{Chusuei_2002}. O and C peaks can often be present in XPS spectra \cite{Chusuei_2002}, a consequence of sample preparation. The rest of the peaks in Fig. SM\ref{XPS} correspond to Ti and Au, confirming the purity of the 1:1 $Pmma$ phase, consistent with neutron (below) and xray (not shown) diffraction data. XPS is also employed in determining the valence of elements in many compounds \cite{Dudy_2013, Kitagawa_1991, Marshall_2011, Novotny_2012}. Fig. SM\ref{XPS}b reveals that the binding energy of $4f_{7/2}$ Au is close to 85 eV, suggesting that Au is close to a Au$^{1+}$ state \cite{McNeillie_1980}. The large energy absorption of Ti does not allow for high resolution measurements, limiting the number and quality of peaks that can be successfully analyzed. In TiAu, the binding energy for the most pronounced Ti $2p_{3/2}$ line is split into two peaks, one at $\approx 455$ eV (for Ti$^{2+}$ or Ti$^{3+}$) \cite{Nist_2012} and another one at $\approx 459$ eV (Ti$^{4+}$) \cite{Nist_2012}, as shown in Fig. SM\ref{XPS}c. A comparison of the areas under the respective curves suggests the valence of Ti to be $3.8 \pm 0.12$. In this itinerant magnet, the XPS results showing a small d electron contribution are consistent with the small itinerant moment ordering indicated by $\mu$SR and neutron measurements (main text), and rule out the presence of a local moment in TiAu.

\section{II. Structural analysis}

\begin{figure}[h!]
\renewcommand{\figurename}{Fig. SM}
\centering
\includegraphics[width=0.75\columnwidth]{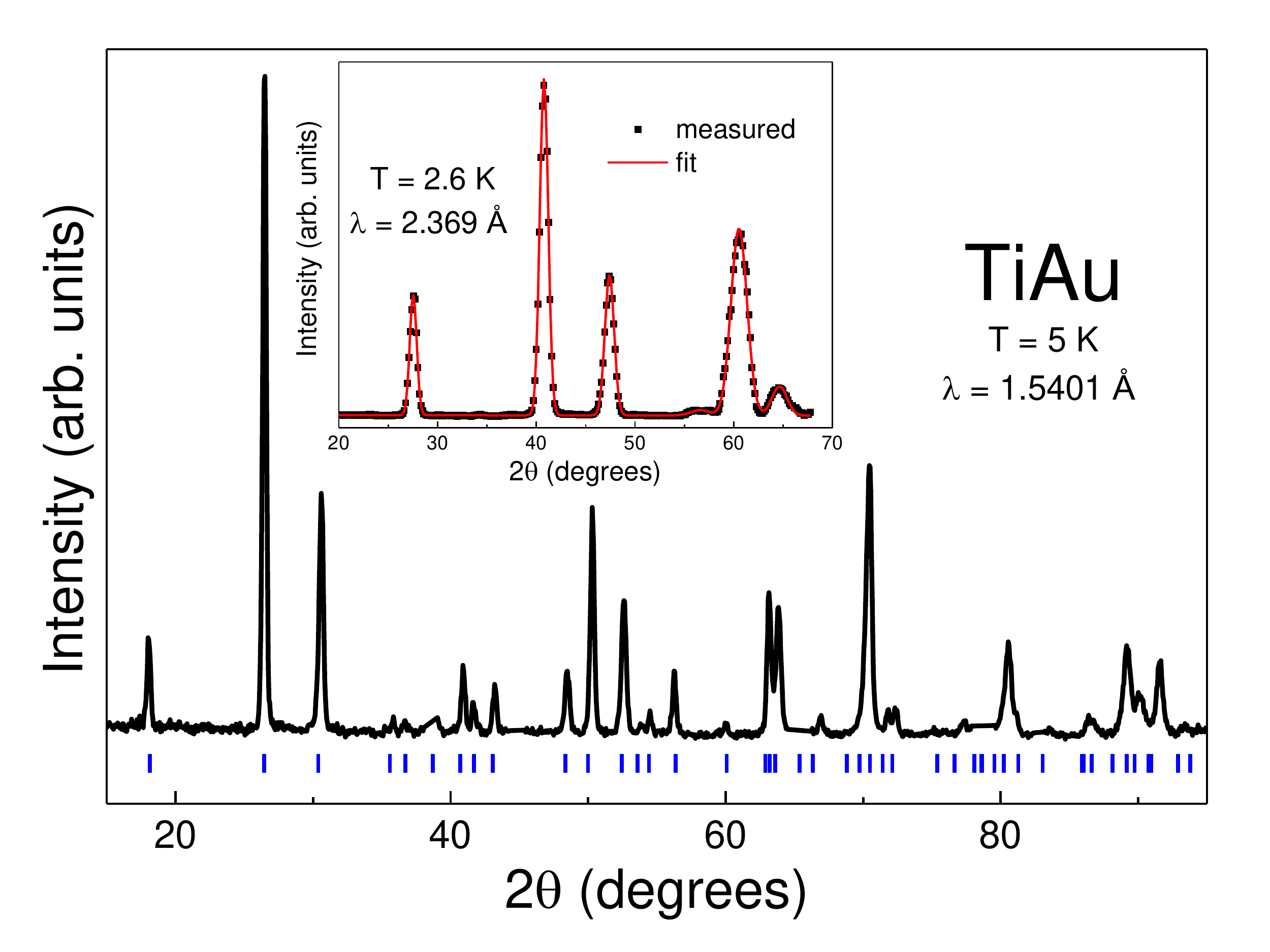}
\caption{\textbf{Neutron diffraction data:} a high resolution diffraction pattern for $T = 5$ K is indexed with the orthorhombic $Pmma$ TiAu phase, denoted by blue vertical marks. Inset: measured T = 2.6 K peaks (squares), measured with high intensity and coarse resolution, together with Gaussian fits (line).}
\label{NIST}
\end{figure}

Neutron diffraction data were collected on the BT-1 powder diffractometer and BT-7 thermal triple-axis spectrometer at the NIST Center for Neutron Research. While BT-1 data were used for the crystallographic analysis, a magnetic peak corresponding to the (0, 1/2, 0) was observed in the BT-7 data (see main text). The BT-1 diffraction pattern recorded at $T = 5$ K (black) is shown in Fig. SM\ref{NIST} with all peaks identified as the orthorhombic $Pmma$ TiAu phase (vertical marks). The inset shows the pattern measured on BT-7 (squares) at $T = 2.6$ K along with Gaussian peak fits (line). The results of structural refinements of the data below ($T = 5$ K) and above ($T = 60$ K) $T_N$ as well as those obtained from room temperature x-ray diffraction are summarized in Table SM II. 

\begin{table}
\renewcommand{\tablename}{Table SM}
\renewcommand{\arraystretch}{1}
\renewcommand{\tabcolsep}{0.2cm}
\caption{\textbf{Crystallographic information for $Pmma$ TiAu}.}
\begin{center}
\begin{tabular}{c|c|c|c} \hline \hline

                & \multicolumn{2}{c|}{Neutron diffraction}             & X-ray diffraction         \\ \hline
Temperature (K) & $T = 5$                  & $T = 60$                 & $T = 300$                 \\ \hline \hline
\multicolumn{4}{c}{Lattice parameters} \\ \hline \hline
a (\AA)         & 4.6221                   & 4.6220                   & 4.6323                    \\ \hline
b (\AA)         & 2.9145                   & 2.9166                   & 2.9489                    \\ \hline
c (\AA)         & 4.8970                   & 4.8959                   & 4.8855                    \\ \hline \hline
\multicolumn{4}{c}{Atomic positions} \\ \hline \hline
Ti              & (0.2500, 0.0000, 0.3110) & (0.2500, 0.0000, 0.3082) & (0.2500, 0.0000, 0.3133)  \\ \hline
Au              & (0.2500, 0.5000, 0.8176) & (0.2500, 0.5000, 0.8176) & (0.2500, 0.5000, 0.8202)  \\ \hline \hline

\end{tabular}
\end{center}
\end{table}

\section{III. Muon spin relaxation}

\begin{figure}
\renewcommand{\tabcolsep}{0.2cm}
\renewcommand{\figurename}{Fig. SM}
\centering
\includegraphics[width=0.75\columnwidth]{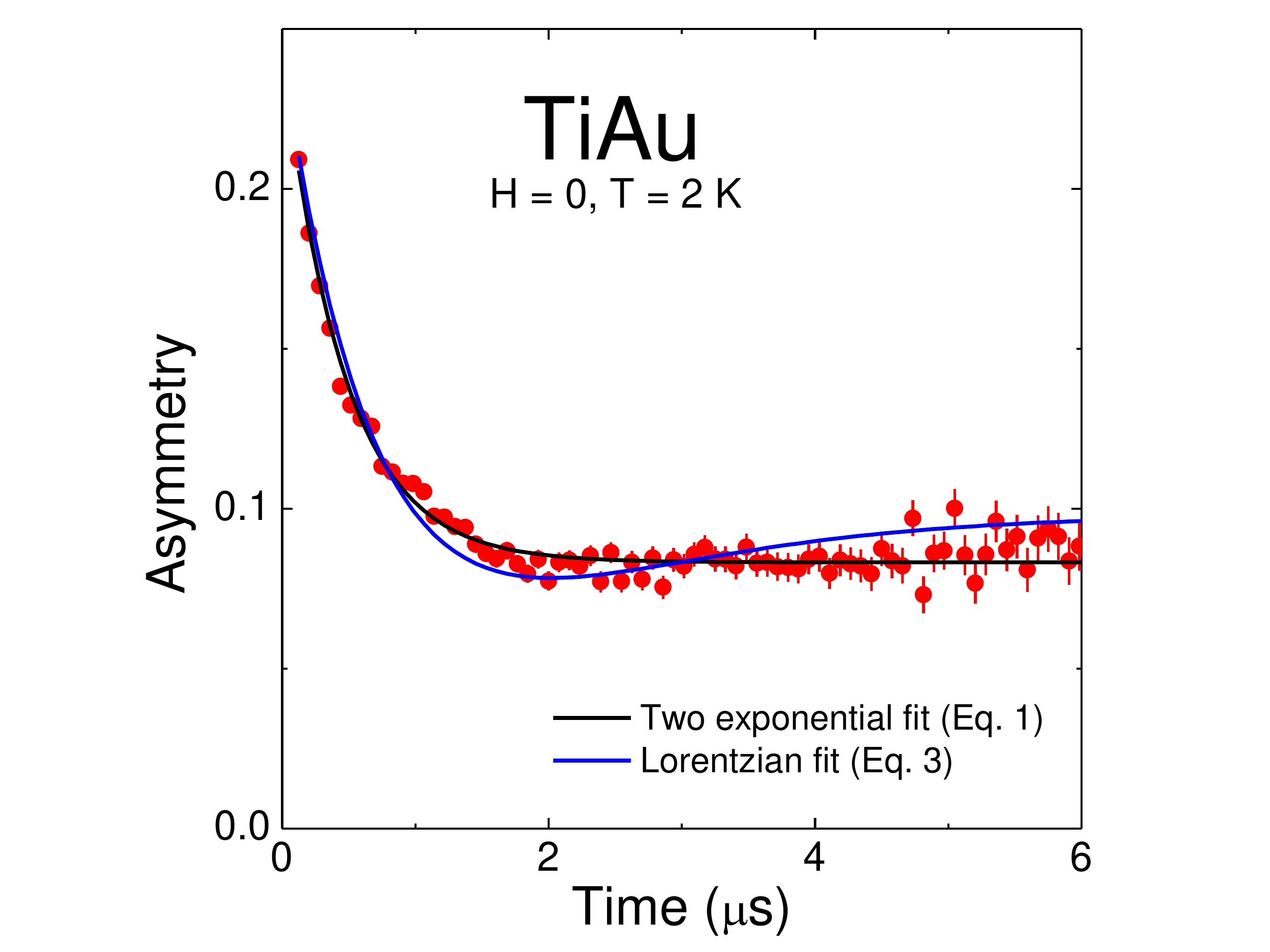}
\caption{\textbf{$\mu$SR assymetry analysis:} the Lorentzian lineshape (blue line) is more appropriate for TiAu rather than the regular two exponential asymmetry expression (black line).}
\label{Fit}
\end{figure}

The initial fit of the muon spin relaxation data included two simple exponential relaxing components, yielding the black curve shown in Fig. SM\ref{Fit}. The asymmetry function had the following form:

\begin{equation}
A(t)=A (f e^{- \lambda_st} + (1 - f) e^{- \lambda_ft})
\end{equation}

However, considering the polycrystalline nature of our sample, the isotropy of overall local magnetic field dictates 1/3 of all muons to have spin parallel to the local field, hence showing no relaxation arising from the random static magnetic field in the sample. In a system with randomly oriented, dense, and static magnetic moments, a Gaussian Kubo-Toyabe relaxation function \cite{Hayano_1979} is usually expected:

\begin{equation}
A_{Gaussian~KT} (t) = A\Big(\frac{1}{3} + \frac{2}{3} (1 - \sigma^2 t^2 ) e^{-\frac{1}{2} \sigma^2 t^2}\Big)
\end{equation}

\noindent while in dilute spin systems a corresponding Lorentzian relaxation function is due to the Lorentzian internal field distribution \cite{Uemura_1985}:

\begin{equation}
A_{Lorentzian~KT} (t) = A\Big(\frac{1}{3} + \frac{2}{3} (1 - at) e^{- at}\Big)
\end{equation}

For the polycrystalline TiAu sample a Lorentzian function is more appropriate and yields a better fit, shown in Fig. SM\ref{Fit} (blue line).

\begin{table}[hbtp]
\renewcommand{\tablename}{Table SM}
\begin{tiny}
\caption{\label{muSRT} \textbf{Comparison of $\mu$SR results for itinerant helimagnetic, FM, SDW, spin glass, and CDW systems.}}
\begin{center}

\begin{tabular}{c|c|c|c|c|c|c|c|c|c|c}
	\hline \hline

                      & Pressure & Chemical             & Relaxation        & Precession  & Ordered& Ground     & Ordered               & Transition  & Dynamical & Reference \\
											&          & composition          & rate ($T = 0$)    & frequency   & volume & state      & moment                & temperature & critical  &           \\
                      & (kbar)   &                      & $\Delta$\footnote{Generally $\Delta$ scales as 2$\pi \nu$} {$\mu$s} & $\nu$ (MHz) & ($T = 0$) (\%) &            & ($\mu_B$ F.U.$^{-1}$) & range (K)   & behavior  &           \\ \hline \hline

TiAu                  & 0        &                      & 1                 &             & 100              & AFM   &                       & 33-38     & no        & Present \\
                  &         &                      &                  &             &               & SDW   &                       &    &         & work \\ \hline
MnSi                  & 0        &                      &                   &  12         & 100              & Helical    & 0.4                   & 29          & yes       & [\citenum{Uemura_2007}] \\ \hline
MnSi                  & 13-15    &                      &                   &  11         & 20-80          & Helical    & 0.4                   & 5, 10       & no        & [\citenum{Uemura_2007}] \\ \hline
SrRuO$_3$             & 0        &                      & 80                &  15, 30     & 100              & FM         & 0.7                   & 160         & yes       & [\citenum{Gat_2011}]    \\ \hline
(Sr, Ca)RuO$_3$       & 0        & Ca$_{0.6 - 0.7}$     & 20                &             & 40-60          & FM         & 0.1-0.2             & 25          & weak      & [\citenum{Gat_2011}]    \\ \hline
(Sr, Ca)$_2$RuO$_4$   & 0        & Ca$_{1.5}$Sr$_{0.5}$ & 8                 &             & 100              & I-SDW      & 0.2-0.3             & 8           & yes       & [\citenum{Carlo_2012}]  \\ \hline
Cu(Mn)                & 0        & 1\% Mn               & 10                &             & 100              & Spin & 0.04& 10          & yes       & [\citenum{Uemura_1985}] \\

                &         &              &               &             &             &glass &(4 per Mn)       &         &       & [\citenum{Uemura_1985}] \\ \hline
BaTi$_2$(As, Sb)$_2$O & 0        &                      &0.1-0.2&             &                  &CDW\footnote{Recently confirmed (unpublished)}&                       &             &           & [\citenum{Nozaki_2013}] \\

&        &                      &(NDB)\footnote{NDB: nuclear dipolar broadening}&             &                  &&                       &             &           & \\ \hline \hline

\end{tabular}
\end{center}
\end{tiny}
\end{table}

\section{IV. Band structure calculations}

\begin{figure}[b!]
\renewcommand{\figurename}{Fig. SM}
\centering
\includegraphics[width=0.75\columnwidth]{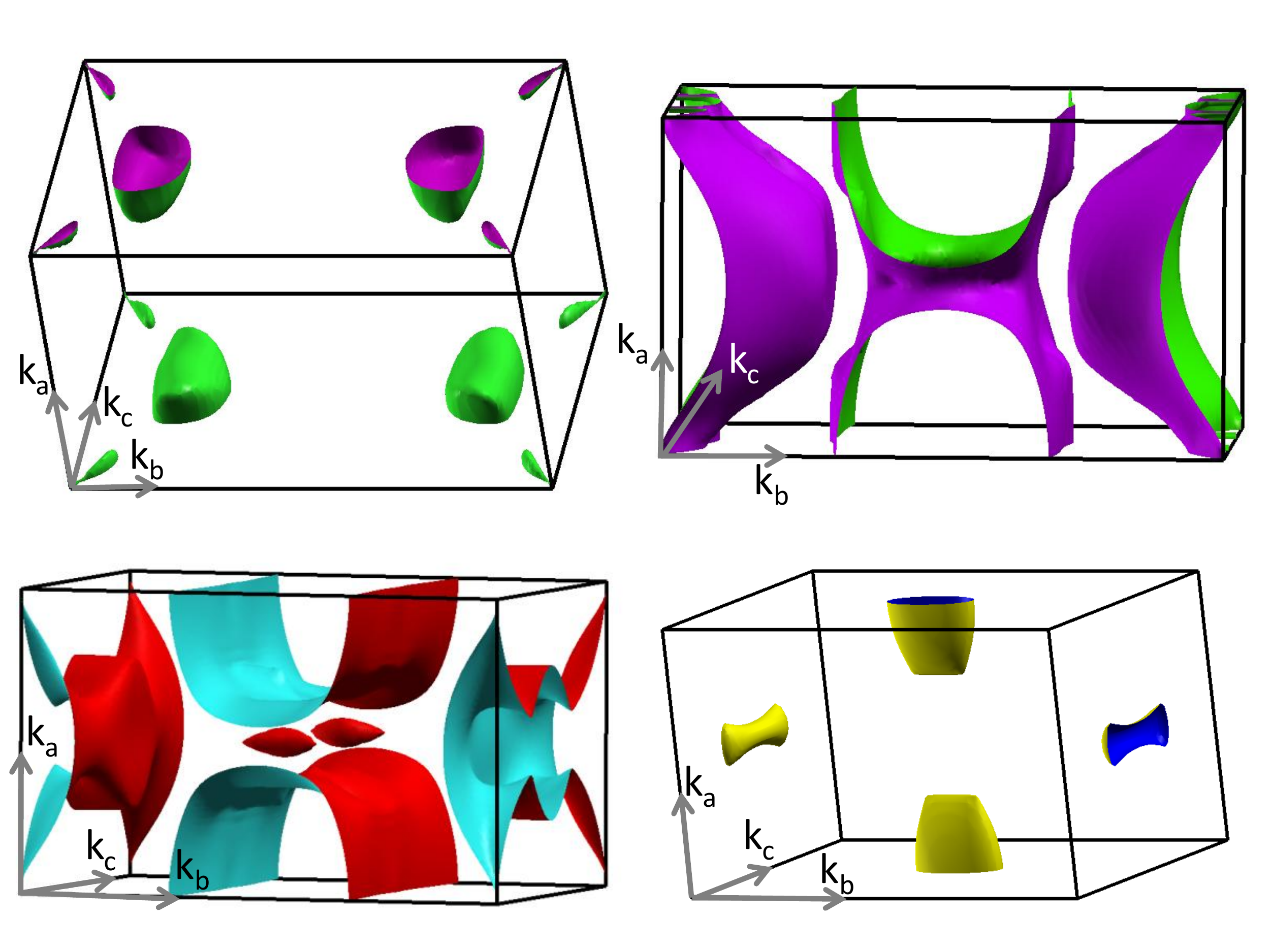}
\caption{\textbf{Separated Fermi surface for different bands.}}
\label{FS}
\end{figure}

Band structure calculations were performed using full-potential linearized augmented plane wave (FP-LAPW) method implemented in the \textit{WIEN2K} package \cite{Blaha}. PBE-GGA was used as the exchange potential, as the default suggestion by \textit{WIEN2K} \cite{Perdew_1996}. A 10$\cdot$10$\cdot$10 k-point grid was used, and shift away from high symmetry directions was allowed. The convergence criterion for force is 1 mRyd/a.u. (1 Ryd = 13.6 eV), with the residual force for the Q = (0, 2b/3$\pi$, 0) state less than 3.5 mRyd/a.u. (or 90 meV/\AA). For the density of states plot, the Gaussian broadening was used, with a broadening factor of 3 mRyd. In order to make the Fermi surface plot (shown in Fig. 5c) easier to read, a separated Fermi surface plot for the different bands in shown in Fig. SM\ref{FS}.

\end{document}